\documentclass[aps,prc,twocolumn]{revtex4}
\usepackage{epsfig}
\usepackage{color}
\begin{document}
\title{Chromoelectric response functions for  quark-gluon plasma}
\author{Akhilesh Ranjan$^1$}
\email{akranjan@iitk.ac.in} 
\author{V. Ravishankar$^{1,2}$}
\email{vravi@iitk.ac.in}
\affiliation{$^1$Department of Physics, Indian Institute of Technology Kanpur,
 UP, India, 208016}
\affiliation{$^2$ Raman Research Institute, Sadashivnagar, Bangalore, 560080
 India}
\date{\today}
\begin{abstract}
We determine the chromoelectric response of quark-gluon plasma 
(QGP) systematically within the framework of classical transport
equations. The transport equations are set up
in the  phase space which includes the $SU(3)$ group space corresponding 
to color (which is a dynamical degree of freedom), in 
addition to the position - momentum variables. The distribution functions 
are defined by projecting the density operators for the quarks and the 
gluons to their respective coherent states (defined over the extended 
phase space). The full import of the Yang-Mills(YM) dynamics is 
shown to manifest through the emergence of an intrinsic  nonlinear, 
nonlocal response, whose behavior we determine in the long 
wavelength limit. It also manifests as a tensor response which is 
a characteristic of gluons. The response functions are shown to have a 
natural interpretation in terms of the renormalizations of the Abelian
and the non-Abelian coupling constants. A detailed analysis of the 
screening of heavy quark potential  and of the exact role played by the 
Debye mass screening in the case of the Cornell potential, is performed. 
We also discuss the non-Abelian contribution to Landau damping in QGP. 
\end{abstract}
\maketitle
\section{Introduction}
The purpose of this paper is to study systematically the color response 
functions of quark gluon plasma, with a proper incorporation of the 
non-Abelian dynamics. Considering  the color electric case, we show the 
emergence of additional response functions which do not have counterparts 
in the well studied electrodynamic plasmas. More precisely, we show the 
emergence of a non-Abelian component of the chromoelectric permittivity in 
the matter and gluonic sectors. The gluonic sector is shown to exhibit yet 
another response, corresponding to its color-tensor exciatations. In this 
work, we merely illustrate these responses in the simple case of ideal 
plasma. We pay particular attention to the role of screening length  
in modifying the dynamics of heavy quarkonium systems. Applications to more 
realistic cases will be taken up in a separate work. The analysis is 
performed within the framework of a classical transport equation -- the anlog 
of Vlasov equation -- as is appropriate to non-Abelian dynamics. We start 
with the motivation below.

Searches for QGP in Ultra Relativistic Heavy Ion Collisions(URHIC) 
have been on experimentally \cite{qgp-def,s-a-bass, exp-status1, exp-status2} 
for more than a decade.  We have today a  wealth of information coming from
observations involving (i) particle multiplicities, (ii) minijets, 
(iii) jet quenching, (iv) strangeness enhancement, (v) collective flows,
(vi) heavy quark dynamics ($J/\Psi$ suppression)\cite{satz},
  (vii) Hanbury Brown-Twiss interferometric measurements\cite{csorgo} etc. 
It is strongly suspected that QGP is already produced in experiments at RHIC.
Further, it has been inferred from flow measurements that the deconfined phase 
is, in all likelihood, a liquid state with a very low viscosity -- 
smaller than even that of liquid helium \cite{liq=he, new-matter}. The upcoming 
experiments at Large Hadron Collider(LHC) may be expected to reveal many more
(unexpected) results.  

A  parallel development that has taken place theoretically is the realization
that QGP exhibits a strong collective behavior which cannot be understood 
perturbatively \cite{lattice,new-matter}. First of all, improved lattice 
computations \cite{lattice,lattice-new} predict that the deconfined phase 
close to the transition temperature $T_c$ is highly non-perturbative, with the
equation of state being far away from that of the ideal gas of quarks and 
gluons. The ideal behavior is not expected to be realized even at $T=2T_c$. 
Analytic studies \cite{g-6} based on improved perturbative approaches (with
computations of $ O(g^6 \rm{log}(g))$ also arrive at the same conclusion
and indicate that the ideal behavior will be seen only at much higher 
temperatures. Experimentally, the flow measurements at RHIC do indicate a low 
viscosity liquid like behavior of the QGP \cite{fluid}. Finally, it is known by
now \cite{pisarski,braaten-1,braaten-2} that a classical behavior emerges 
naturally when one considers hard thermal loop(HTL) contributions. A local 
formulation of HTL effective action has been obtained by Blaizot and Iancu who
have succeeded in rewriting the HTL effective theory as a kinetic theory with
a Vlasov term \cite{blaizot-1, blaizot-2, blaizot-3}. 

It may be, therefore, a fruitful endeavor to take the classical frame work 
seriously and explore the extent to which it can capture the expected features
of QGP. It is well recognized that its success would depend crucially on the 
availability of reasonable mechanisms for (i) the production of soft partons
(rate term in the phase space) and (ii) the expression for the equation of 
state and its equilibrium distribution function. Indeed, significant 
progress has 
been made in both the directions by employing a variety of techniques 
\cite{geiger, lary,ravi, raju}, although they are all not always 
mutually consistent. Be it as may, we wish to point out in this paper that 
the evaluation of the response functions of the QGP, so crucial for the 
signals, is further dependent on the form of the distribution functions in
the color space, which is inherited from the parent 
density operators; it may not be assigned at will. 

The central theme of the paper is to couple the coherent state representation
in the color space with the YM dynamics in the transport equations. In doing 
so, we derive, in  a consistent manner,  the response functions which show 
manifestly the Abelian and the non-Abelian  contributions. The latter is 
nonlinear and nonlocal in space time. Hence it is also non-Markovian. We also 
show the intrinsic distinction between a pure QCD plasma which is purely 
gluonic and QGP which has both matter and gluonic components. 
 Since we do not wish to make any realistic calculation 
here, many simplifications will be made in the transport equations: (i) the 
equilibrium distribution functions will be taken to be ideal, (ii) we study 
the response when the system is close to equilibrium, so that the 
source term may also be dropped and (iii) finally, we take the plasma to be 
spatially isotropic. The plasma is taken to be collisionless. 
These simplifications will be improved upon 
in a separate paper, where more realistic collision terms such as the one 
derived by Bodeker \cite{bodeker} will be used. As remarked above, the 
primary purpose of the paper is to elucidate the structure of the response  
functions and to illustrate it in simple situations. 

The chromoelectric response functions are particularly important, ever since 
Matsui and Satz \cite{matsui} predicted  $J/\Psi$ suppression as a signature 
for QGP in heavy ion collisions. Lattice calculations \cite{lattice} do 
support the prediction;  early estimates of the suppression in heavy ion 
collisions have involved the assumption of a hydrodynamic expansion of an 
ideal plasma. The suppression is of course determined by the Debye mass which 
evolves with the temperature.  We study the precise role of the Debye mass in 
the specific case of the Cornell potential. In any case, it appears that the 
Debye mass is an incomplete manifestation of the non-Abelian dynamics, in as 
much as that it {\it does not} require the non-Abelian interactions for its 
emergence. It is not unreasonable to look for exclusively non-Abelian 
responses which would be nonlinear, and reflect additional properties of 
chromodynamics in a medium. We do show that such a non-Abelian permittivity 
exists. The new permittivity may also lead to other signatures which can 
perhaps be tested experimentally.

\subsection{A brief review} 

  We briefly review the studies on the QGP response functions so far to the 
extent that they are relevant to our work.
 There have been many attempts to calculate the response function for QGP. 
An approach based on perturbative QCD (pQCD) at finite temperature was
adopted by Weldon \cite{weldon} who studied plasma screening and plasma 
oscillations. Based on an Abelian analysis, Mustafa, Thoma and Chakraborty
 \cite{thoma} have argued for bound quark structures, which have also been 
predicted by Liao and Shuryak
\cite{shuryak}. The conclusion is essentially based on their 
analysis for moving partons. Petreczky \cite{peter} has, on the other hand, 
considered a $SU(2)$ plasma, and extracted the chromoelectric screening by
 analyzing the long distance behavior of the static part of the longitudinal 
propagator in the lowest order. His results are off by about 25$\%$ from 
lattice simulations, and are not easily adapted
 to the scenario prevalent in URHIC. Finally, in a work somewhat close to ours,
 Chen {\it et al} \cite{chen} have employed the Vlasov equation to study the 
color response.
However, the gauge invariance
 of their results is not apparent, as also its applicability to URHIC. None of 
the above approaches employs the coherent state projections which establish a 
natural connection between the underlying quantum dynamics with the classical 
approximations. They do not also study exhaustively {\it all possible}
response functions involving the color degree of freedom.

\section{Determination of the response functions}
As mentioned above, the approach will be based on the transport equation for 
the phase space distribution functions for the quarks and the gluons. The 
distribution functions for an $N$ particle system will have the generic form 
$f(\vec{q}_1, \cdots \vec{q_N};~~ \vec{p}_1, \cdots \vec{p}_N; ~~Q_1^a, \cdots
Q_N^a)$ in terms of the coordinates $\vec{q}$, the momenta $\vec{p}$, and the 
color charges $Q^a$ which are defined in the space corresponding to the 
gauge group. Note that the distribution function is invariant under  gauge 
transformations (for a formal demonstration, see {\it e.g.}, \cite{cristina}). 
 Let us first consider the generic form of the transport equation

\begin{eqnarray}
\label{eq1}
\frac{\partial f}{\partial t}+v_i 
\frac{\partial f}{\partial x_i}+F_i \frac{\partial f}{\partial p_i}+\dot{Q^a} 
\frac{\partial f}{\partial Q^a}=\Sigma+C, 
\end{eqnarray}
where the source term is denoted by  $\Sigma$ and $C$ is the 
collision term.  
The source term is indeed important in studying the production and evolution 
of QGP \cite{bhal,ravishankar,schmidt}. But as mentioned above, we study the 
response when system is close to its equilibrium configuration, whence the 
source term may be dropped. We further work in the Vlasov limit so that the 
collision term will also be dropped. The responses are, of course eventually 
evaluated in the limit that the collision time $\tau_c \rightarrow \infty$, 
via the standard Landau prescription, The reason for working in this limit
is that the inclusion of any realistic collision term, as, for instance, the 
one derived by Bodeker \cite{bodeker} will complicate the discussions. On the 
other hand, a simple minded relaxation term would violate current 
conservation \cite{gross}. This draw back will be remedied in a separate 
paper.

The third term in Eq.$\ref{eq1}$ represents the Vlasov term corresponding 
to the action of the mean field on the plasma. The next term, involving the 
derivative of the color charge, is unique to the non-Abelian plasma, and 
gives the dynamical variation of the color charge (in the compact part of 
the phase space). The dynamics modulates and modifies the contribution of 
the standard Vlasov term; further, it also gives rise to the non-Abelian 
component of the response function.  Our analysis makes crucial
use of the most general form of the quark and the gluonic distribution 
functions in the color variables, and we take up the determination of their 
form in the next section.

\subsection{The distribution functions}
\subsubsection{The coherent bases}
 We elucidate the procedure for obtaining the generic forms 
of phase space distribution functions from their underlying quantum states. 
The best way of extracting them is to project the parent quantum state to a 
coherent basis. Indeed, with the usual position-momentum variables, the 
coherent states are the closest to the classical states since they 
possess  minimum uncertainty in position and momentum. The projection  
defines, in a natural manner, the area over which the phase space is coarse 
grained: it is simply given by $2\pi \hbar$ for each degree of freedom. 
Further, the coherent state projection is equivalent to the Wigner 
distribution function used {\it e.g.} by Elze and Heinz \cite{elze,heinz}. 
However, it has the advantage that unlike the Wigner distribution function 
which can take negative values, the phase space density  obtained from the 
coherent state is always non-negative.

The above method needs a refinement when it comes to the color degree of 
freedom. Recall that the classical phase spaces associated with finite 
dimensional Hilbert spaces are compact (as for example, in the case of spin). 
There is no neat separation of  canonically conjugate
variables, and there is no associated commutator (of the kind $[x,p] =i$) 
that leads to minimum uncertainty states. Instead,
the commutators  are given by the Lie Algebra of the associated compact group. 
Care needs to be taken in defining the phase space distribution in this case. 
The most convenient method is to follow Perelomov \cite{perel-1}:  take 
a representative state, and act the group $exp\{ i T_a \theta_a \}$ on a 
reference state which is typically taken to be the state with the highest 
weight. The resulting orbit, ${\cal O} \big (\vert \{\theta_a \}> \big)$ 
forms a faithful copy of the associated phase space. The quantity
$\langle \theta_1,\cdots \theta_n \vert \hat{\rho}  \vert \theta_1,\cdots \theta_n  \rangle$ is interpreted consistently as the classical probability 
density in the color part od the phase space. These distributions so obtained are always 
smeared, for any finite dimensional representation of the compact group.

\subsubsection{Distribution functions for quarks and gluons}
We employ the method outlined above for the specific system of interest.
 Let $\hat{\rho}$  be the density operator for a parton.  Denote by 
$\vert \Psi >_c = \vert \vec{r},~\vec{p} >\otimes \vert Q^a >$ the coherent 
basis for the parton, where the first term is the usual minimum uncertainty 
state. As mentioned, the latter term is obtained by the action of the gauge 
group on a standard state, $\vert \psi>_s$,  taken to be the one with the 
highest weight \cite{perel-1}. Thus the coherent states in the color sector 
have the form $\vert Q^a > = exp \{i Q^a T^a \} \vert \psi_s>$, where the 
variables $Q^a$ provide a coordinate description of the group space. For 
example, the parameters for $SU(2)$ may be chosen to be the Euler angles. A 
generalized Euler angle description for $SU(3)$, which is of relevance to us 
here has been provided by Byrd \cite{byrd}. We will be using it subsequently.

The form of the generators $T^a$ are to be chosen depending on the parton, 
{\it i.e.}, the representation to which it belongs. The quarks transform 
according to the fundamental representation of the gauge group and the 
generators will also be chosen accordingly.  Gluons, on the 
other hand, belong to the adjoint representation. As a direct consequence, 
the dimension of the phase space will be six for the quarks, and eight 
for the gluons if we consider $SU(3)$ gauge group. Incidentally, note 
that the phase space for the quark is isomorphic to its Hilbert space. The 
gluonic phase space constitutes, in contrast, a coset space, being an orbit 
of the gauge group in the Hilbert space which is of dimension fifteen. The 
property of the coherent state depends naturally on the orbit to 
which it belongs. Suffice to say, the color coherent states are natural 
generalizations of the more well known spin coherent states. We refer the 
reader to literature \cite{perel-1} for more details.

The structure of the copy of the phase space, obtained thus as a coset space 
has strong implications on the nature of the color distributions, bearing 
a deep connection with the classical-quantum correspondence. For any finite 
dimensional representation of the group, only a finite number of the moments 
of the variables can contribute. To illustrate with a simple example, the 
state of a spin half particle is characterized entirely by the vector 
polarization Tr$\{\rho \vec{S}\}$. A spin one state requires, in addition, a 
specification of the second order tensor polarization 
Tr$\{\rho S_{ij}\}$, where $S_{ij}=\frac{S_iS_j +S_jS_i}{2} -\frac{1}{3}{\vec{S}}^2 \delta_{ij}$. 

We perform a similar analysis for quarks and gluons. Since the quarks belong 
to the fundamental representation of the gauge group, 
the quark distributions can at the most be linear in the color charge $Q$. All 
the higher order terms vanish identically.
Thus, the single quark distribution function has the most general form
$ f = f_0 + f^a \hat{Q}^a$ where $f_{0},~f^a$ are functions of the usual 
phase space variables $\{x_i,~p_i\}$. In contrast, the gluons belong to the 
adjoint representation; they admit a more general expansion, 
with an additional bilinear term in the color charge. Note that the Abelian 
limit is obtained by dropping all the multipole terms except the scalar 
component. The classical limit is obtained in the other limit, the so called 
large $N_c$ limit, by admitting multipole contributions of arbitrary high 
orders. Finally, it may be noted that the two Casimirs $Q^a Q^a$ and 
$d^{abc}Q^aQ^bQ^c$ restrict the dynamics under the gauge independent 
interactions to a smaller subspace. As we shall see latter, the above 
resolution of the distribution functions brings out the richness in the 
response functions.

\section{The response functions}

\subsection{The quark-antiquark sector}
Strictly speaking, one should write a two particle ditribution function for 
the $q\bar{q}$ system to study the color charge excitations. This procedure 
is laborious \cite{arxiv} and we present a simplified version below. Consider 
color deviation of the $q\bar{q}$ system from its equilibrium configuration 
$f_0$, which is taken to be a uniform distribution in color space. 
Accordingly, we write the distribution function in the form
\begin{equation}
\label{eq3}
f=f_0 (p)+\hat{Q}^{a}f^{a}(x_i,p_i),
\end{equation}  
  which emphasizes that
$f_0$ is independent of position and color variables. The color fluctuations
which are obtained by perturbing around $f_0$ are denoted by 
$ \hat{Q}^{a}f^{a}$. The color density functions $f^{a}$ are functions of 
both position and momentum. Finally, we also note that at equilibrium, the 
chromoelectric field vanishes identically.
  
 The transport equation reads

\begin{eqnarray}
\label{eq4}
 \frac{\partial f}{\partial t}+v_{i}\frac{\partial f}{\partial x_{i}}
 +F_{i }\frac{\partial f}{\partial p_{i}} +\dot{Q^{a}}\frac{\partial f}
 {\partial Q^{a}}=0,
\end{eqnarray}
 which on the application of the Wong equation \cite{wong}
\begin{equation}
\label{eq5}
\frac{dQ^a}{d\tau}=f^{abc}Q^bu_{\mu}A^{\mu c},
\end{equation}         
 acquires the form
\begin{eqnarray}
\label{eq6}
&&\frac{\partial f}{\partial t}+v_{i}\frac{\partial f}
{\partial x_{i}}+Q^{a}E^{a}_{i}(\vec{r})
\frac{\partial f}{\partial p_{i}} \nonumber
 \\
&&-f^{lmn}\{A^l_0(\vec{r})-v_{i}A^l_i(\vec{r})\}  
 Q^{m}\frac{\partial f}
 {\partial Q^{n}}
=0,
\end{eqnarray} 
in obtaining which we have assumed that the plasma is spatially isotropic 
and that the chromomagnetic field is absent. Unlike the electrodynamic 
plasma, the above equaton needs a specification of the gauge\cite{Wu} 
\footnote{ Unlike as in electrodynamics, the gauge invariant content of 
the field is not exhausted by just giving the field tensor. In fact, the 
same field tensor can be obtained by inequivalent gauge potentials 
\cite{Wu}. Hence the gauge potential in Wong equation Eq.4 has nontrivial 
consequences.}. We study Eq.$\ref{eq6}$ in the temporal gauge, 
 $A^a_0=0$. In this gauge the chromoelectric field has a 
simple dependence on the gauge potential:
$$E^a_i(\vec{r},t)=-\frac{\partial{A}^a_i(t,\vec{r})}{\partial t}.
$$
The transport equation  now gets a simpler form
\begin{eqnarray}
\label{eq7}
&&\frac{\partial f}{\partial t}+v_{i}\frac{\partial f}
{\partial x_{i}}+Q^{a}E^{a}_{i}(\vec{r})
\frac{\partial f}{\partial p_{i}}+f^{lmn}v_{i}A^l_i(\vec{r})  
 Q^{m}\frac{\partial f}
 {\partial Q^{n}} =0.
\end{eqnarray} 

 The evaluation of the chromoelectric response requires the determination of 
the equations for the moments of the distribution function, involving 
integrations in the group space by employing the appropriate Haar 
measure in the group space.
The details of the evaluation of the color moments are provided in the 
Appendix.  

At this juncture, it
is convenient to switch over to the vector notation for color variables: 
$$
f^aQ^a \equiv \vec{f} \cdot \vec{Q};~~f^{lmn} A^l B^m \equiv (\vec{A} \times \vec{B})^n.
$$  
The latter notation collapses to the usual notation when the gauge group is 
$SU(2)$. We continue to denote the position and momenta by their components. 
Thus, we write
\begin{equation}
\label{impeq}
\frac{\partial \vec{f}}{\partial t} +v_i\frac{\partial \vec{f}}{\partial_{x_i}} +Q \vec{E}_i \frac{\partial f_0}{\partial_{p_i}}
 -v_i\vec{A}_i \times \vec{f} =0.
\end{equation}
Note that in obtaining the above equation we have further assumed that $\vec{f} $ is a small perturbation on $f_0$. In particular, 
$\frac{\partial f_0}{\partial_{p_i}} $ dominates over the gradient of the 
color vector fluctuation  $\vert\frac{\partial \vec{f}}{\partial_{p_i}}\vert$ which 
we drop.

The color response function is to be derived from Eq.\ref{impeq} which, as 
we see, simply replaces the ordinary derivative by the covariant derivative, 
in the temporal gauge. It turns out that the extraction of the response
function is not as straight forward as in the case of ordinary plasma.
We follow the standard treatment and write Eq.\ref{impeq} in the Fourier 
space. We denote by $\vec{\phi}(\omega,k_i, p_i)$, the Fourier transform 
of $\vec{f}(t, x_i, p_i)$. Similarly, $\vec{{\cal E}}_i$ and 
$\vec{{\cal A}}_i$ represent the Fourier transforms of the gauge field and 
the gauge potential respectively. We obtain

\begin{equation}
\label{fteq}
\vec{\phi}(\omega,k_i,p_i) = -Q \vec{{\cal E}}_i \frac{1}{{\cal D}} \frac{\partial f_0}{\partial_{p_i}} +
 (\vec{{\cal A}}_i \star\vec{\phi})(\omega,k_i,p_i)\frac{1}{{\cal D}} v_i.
\end{equation}
In the above expression, ${\cal D} = i\omega -ik_iv_i$.
We have employed the notation $\star$ to denote the convolution of the 
Fourier transforms, combined with the cross product in the color variables:
$$
(\vec{{\cal A}} \star \vec{{\cal B}})^c  \equiv f^{abc} \int d\omega^{\prime} d^3 k^{\prime} A^i(\omega -\omega^{\prime}, k_i-k_i^{\prime})B^b(\omega^{\prime}, k_i^{\prime}).
$$
 Q denotes the magnitude of the color charge.

The expression for the charge density $\vec{\rho}$ is easily obtained to be
\begin{equation}
\label{rho}
\vec{\rho}(\omega,k_i) = -Q^2 \vec{{\cal E}}_i \int d^3p \frac{1}{{\cal D}} \frac{\partial f_0}{\partial_{p_i}} 
 +Q(\vec{{\cal A}}_i \star \vec{\Phi}_i)(\omega, k_i),
\end{equation}
where we have defined
\begin{equation}
\label{defPhi}
\vec{\Phi}_i(\omega, k_i) =\int d^3p v_i\frac{1}{{\cal D}}\vec{\phi}(\omega,k_i,p_i) .
\end{equation}

It is clear from the above equation that the expression for $\vec{\rho}$ is 
not closed, even after we employ the Gauss Law expression, all beacuse of the 
second term which cannot be written in terms of the charge density or the 
field. A rigorous evaluation would lead to an infinite hierarchy of equations
with moments of the distribution functions with respect to velocities. Since 
we are interested in the bulk properties of the medium, we evaluate 
Eq.\ref{rho} in the long wavelength limit, by assuming that 
$k_i /\omega \ll 1$. In this approximation, Eq. \ref{defPhi} simply becomes 
$$
\vec{\Phi}_i(\omega, k_i) = \frac{1}{\omega} \vec{J}_i,
$$
in terms of the color current density $\vec{J}_i$. It follows from Eq.8 that 
the equation for $\vec{J}_i$ is given by 
\begin{equation}
\label{current}
\vec{J}_i(\omega,k_i) = -Q^2 \vec{{\cal E}}_i \int d^3p \frac{1}{{\cal D}} v_i\frac{\partial f_0}{\partial_{p_i}} 
 +Q(\vec{{\cal A}}_i \star \vec{\Phi}_{ij})(\omega, k_i),
\end{equation}
where 
$$
\vec{\Phi}_{ij}(\omega, k_i) = \int d^3p v_iv_j\frac{1}{{\cal D}}\vec{\phi}(\omega,k_i,p_i),
$$
which again involves an expression which is one order higher in the moment 
with respect to velocity. We truncate the hierarchy, by dropping the 
contribution from the second term, and write 
\begin{equation}
\label{currentfin}
\vec{J}_i = -Q^2 \vec{{\cal E}}_j \int d^3p v_i\frac{1}{{\cal D}} \frac{\partial f_0}{\partial_{p_j}}.
\end{equation}

We may thus rewrite Eq.\ref{rho} in the form 
\begin{equation}
\label{perm1}
\vec{\rho}(\omega,k_i,p_i) = -iQ^2 \vec{{\cal E}}_i I_i
-Q^2 \frac{1}{\omega}\vec{{\cal A}}_i \star (\vec{{\cal E}}_j I_{ij}).
\end{equation}
The integrals $I_i, I_{ij}$ are entirely functions of the equilibrium 
distribution functions, and given by
\begin{eqnarray}
I_i & = \sqrt{-1}& \int d^3p \frac{1}{{\cal D}} \frac{\partial f_0}{\partial_{p_i}}, \nonumber \\
I_{ij} & = \sqrt{-1}& \int d^3p v_i\frac{1}{{\cal D}} \frac{\partial f_0}{\partial_{p_j}}.
\end{eqnarray}
The determination of  permittivity is almost complete. Since we are 
considering an isotropic system, we observe that the integrals above have 
the resolutions
\begin{eqnarray}
 I_i &= & k_iI_0(\omega, k) \nonumber \\
 I_{ij} & = & \delta_{ij}I_1(\omega,k)+k_ik_jI_2(\omega,k).
\end{eqnarray}
Clearly, $I_2 $ is analytic in $k$ and, thus does not contribute at low 
wavelengths. The final expression is thus given by
\begin{eqnarray}
\label{Perm}
\vec{\rho}(\omega,k_i)+iQ^2 k_i\vec{{\cal E}}_i(\omega,k_i)I_0(\omega,k)
-\frac{Q^2}{\omega} \vec{{\cal A}}_i \star (\vec{ {\cal E}}_iI_1) =0.
\end{eqnarray}
The response functions can be inferred by comparing Eq.\ref{perm1} with the 
Gauss' law in vacuum in the Fourier space: 
\begin{eqnarray}
\label{eq11}
\vec{\rho}(\omega,k_i)-ik_i\vec{{\cal E}}_i(\omega,k_i)+
\vec{{\cal A}}_i \star \vec{{\cal E}}_i =0.
\end{eqnarray}
Combining Eqs.\ref{perm1} and \ref{eq11}, we are led to define the following 
permittivities
\begin{eqnarray}
\label{eq12}
\epsilon^{ab}_{ij}(\omega,k)=\{1+Q^2I_0(\omega,k)\}\delta_{ij}\delta^{ab}
\equiv \epsilon_A\delta_{ij}\delta{ab}.
\end{eqnarray}

\begin{eqnarray}
\epsilon^{abc}_{ij}(\omega,\omega')=\{1+\frac{Q^2\left.I_1(\omega',k^{\prime})\right|_{k^{\prime}=0}}{\omega}\}f^{abc}\delta_{ij}
\nonumber
\end{eqnarray}
\begin{eqnarray}
\label{eq13}
\equiv\epsilon_N f^{abc} \delta_{ij}.
\end{eqnarray}

Note the emergence of the new permittivity $\epsilon_N$ which has no
 Abelian counterpart. It is an invariant  tensor of rank three in the color 
space (being proportional to $f^{abc}$). The new permittivity emerges, we 
stress,  in addition to the Abelian permittivity $\epsilon_A$, which 
leads to the standard screening. As much as the Abelian permittivity signifies 
a renormalization of the charge, the non-Abelian permittivity exhibits the 
renormalization of the non-Abelian coupling coefficients $f^{abc}$. Gauge 
invariance of the theory constrains the two renormalizations, as shown in 
the above equations, via the dependence on $f_{eq}$. Explicitly, the two 
integrals $I_0$ and $I_1$ which define the two permittivities may be generated 
from a single function:
\begin{eqnarray}
I_0= \frac{1}{k^2}\frac{\partial}{\partial \omega} \int ln({\cal D})k_i \partial_{{p_i}}f_{eq} d^3p \\
I_1= -\frac{1}{k^2}\frac{\partial}{\partial k_i} \int ln({\cal D}) \partial_{p_i}f_{eq} d^3p.
\end{eqnarray}
Note that all the ``couplings" $f^{abc}$ undergo the 
same renormalization, as required by gauge invariance.
 
The resolution of the permittivity into the Abelian and the non-Abelian 
sectors is itself dependent on the gauge employed. The above identification 
is done in the temporal gauge. However, the results are themselves gauge 
invariant. We refrain from writing the response functions and the counter 
parts of Eq.\ref{eq13} in an arbitrary gauge as it is rather cumbersome, 
and of no use to us here. 

\subsection{Form of the permittivities}
It is instructive to study the form of the response functions in the 
simple case when $f_{eq}$ has the standard Fermi-Dirac form. The expression 
for the Abelian permittivity reads 
\begin{eqnarray}
\label{eq14}
\epsilon_A=1+\frac{2\pi^3Q^2T^2N_f}{3k^2}\{-\frac{\omega}{k}ln\left|\frac{\omega+k}{\omega-k}\right|+2\}
\end{eqnarray}
and is not different from that of the electrodynamic plasma, except for  
multiplicative color and flavor factors. The non-Abelian permittivity is 
however, novel, and has the form

\begin{eqnarray}
\label{eq15}
\epsilon_N
=\big\{1-\frac{4\pi^3Q^2T^2N_f}{9}\frac{1}{\omega\omega'}\big\}.
\end{eqnarray}
Although the functions are real, their imaginary components can be extracted 
by employing the standard Landau $i\epsilon$ prescription. The imaginary 
components cause the Landau damping which we will discuss in detail in 
a separate section.

\subsection{Physical significance of $\epsilon_N$}
The significance of $\epsilon_N$ is best seen
in the induced non-Abelian component of the charge density, the so called 
charge density carried by the field. 
In the Gauss' equation for non-Abelian fields the term $\rho^f_a\equiv f^{abc}
A^b_i E^c_i$ (where superscript `$f$' stands for field component) has a natural
interpretation of the charge density contributed by the field. This charge 
density is not gauge covariant. However, the total charge is gauge covariant, 
provided reasonable boundary conditions are imposed. $\epsilon_N$ may 
now be looked upon as modifying $\rho^f_a$ by inducing additional charges. 
To study this we evaluate the modification to $\rho^f_a$
and obtain
\begin{eqnarray}
\label{eq16}
\rho^f_a=-f^{abc}A^b_i(t,\vec{r})E^c_i(t,\vec{r})+\frac{\pi Q^2T^2N_f}{9}f^{abc}\nonumber\\
\times\int_{-\infty}^t \,dt' A_i^b(t',\vec{r})\int_{-\infty}^{t'} \,dt'' E_i^c(t'',\vec{r}),
\end{eqnarray}
where the second term shows the induced charge density of the field.
 The above equation displays the inherently nonlinear, non-Markovian nature
of the non-Abelian  permittivity. In the collisionless limit that we are 
interested, the response is maximally non-Markovian.

It is pertinent at this stage to mention that the non-Abelian response plays 
an important role involving three gluon processes, especially gluonic 
bremsstrahlung of gluons, and the analog of \v{C}erenkov radiation of gluons. 
Their phenomenological significance in heavy ion collisions remains to be 
investigated.

\subsection{ The gluonic sector}

  We now consider the bosonic content of plasma. It differs from the matter 
sector in two respects. Its equilibrium distribution function is given by 
the Bose-Einstein form, and it has a richer structure in the color space. 
As a warm up, we first consider $SU(2)$, which is simpler.

\subsubsection{ $SU(2)$ gluons}

The color coherent states are no different from the more familiar spin coherent
states (corresponding to spin -$1$) in the adjoint representation. The density 
operator in the coherent basis has the expansion

\begin{equation}
\label{eq17}
f(\vec{r},\vec{p},Q^a)=f_0+\hat{Q}^{a}f^{a}+\hat{Q}^{ab}f^{ab},
\end{equation} 
where the equilibrium distribution $f_0$  corresponds to the singlet, 
$f^{a}$ to the triplet $D^1$, and $f^{ab}$ to the 5-dimensional irreducible 
representation $D^2$ of $SU(2)$. Recall that $f^{ab}$ is a completely 
symmetric and traceless matrix, as is also $\hat{Q}^{ab}$ which is given by,
\[Q^2 \hat{Q}^{ab}=Q^aQ^b-\frac{\delta^{ab}}{3}Q^2. \]
There are, in all, nine independent functions of position and momentum. Of 
them, the tensor components $f^{ab}$ are specific to the gluonic sector and 
are absent in the quark sector. Thus the perturbation around the equilibrium 
configuration has a richer structure for the gluons than for the quarks in the 
color space. We  treat $f^a$ and $f^{ab}$ to be 
fluctuations around $f_0$.

We may now repeat the evaluation of the moments in the color space as in the 
quark sector. Since the tensors are irreducible, the quark sector results hold 
in the gluonic sector as well for the charge density leading essentially to the
 same expression for the permittivities

\begin{eqnarray}
\label{eq17a}
\epsilon_A=1+\frac{4\pi^3Q^2T^2}{3k^2}\{-\frac{\omega}{k}ln\left|\frac{\omega+k}{\omega-k}\right|+2\},
\end{eqnarray}

\begin{eqnarray}
\label{eq17b}
\epsilon_N
=1-\frac{8\pi^3Q^2T^2}{9}\frac{1}{\omega\omega'}.
\end{eqnarray}
These differ from the expressions for quarks because of the choice of $f_0$. 
However, unlike in the case of the quarks, the above permittivities {\it do not}
 capture completely the response of the medium. There are additional 
contributions coming from the tensor fluctuations $f^{ab}$. To study them, we
evaluate the moment of the transport equation {\it wrt} $Q^{ab}$. 
In the spirit of the earlier calculations, we keep only the contribution from 
$f_0$ in the terms involving the gradient {\it wrt} the momentum. In this 
case, it is preferable to evaluate the moment in the phase space, {\it i.e.}, 
without integration over momentum. Let us denote by 
$\phi^{ab}(\omega, k_i, p_i)$ the Fourier transform of $f^{ab}$. We then get 
\begin{eqnarray}
\label{eq18}
\phi^{ab}( \omega, k_i, p_i)-\frac{i}{\omega} L^{abcde}v_i({\cal A}^c_i \otimes \phi^{de})
( \omega, k_i, p_i) =0,
\end{eqnarray}
where the symbol $\otimes$ denote only the convolution of the Fourier 
transforms, without any operation in  the color space. The fifth rank 
invariant tensor is given by
\begin{equation}
L^{abcde} = i(\epsilon^{acd}\delta^{be}+\epsilon^{bcd}\delta^{ae}).
\end{equation}
The above expression is an eigenvalue equation in the tensor fluctuations 
$f^{ab}$, in the extended phase space. The spectrum is, significantly, 
independent of the equation of state, {\it i.e.}, of $f_0$ --  within the 
approximate framework employed here. It is determined entirely by the structure
constants and, of course, the perturbing gauge fields. It should be of interest 
to work out the experimental consequences, in terms of the propagation of 
gluons in the medium.

To gain further insight, we integrate Eq.\ref{eq18} over the momentum 
variables. We obtain
\begin{eqnarray}
\label{eq19}
\rho^{ab}( \omega, k_i)-\frac{i}{\omega} L^{abcde}({\cal A}^c_i \otimes J_i^{de})
( \omega, k_i) =0,
\end{eqnarray}

which establishes a relation betwen the tensor charge density
$$
\int Q^2\phi^{ab} \,d^3\vec{p}= {\rho}^{ab},
$$
with the tensor current density
$$
\int Q^2 v_i \phi^{ab} \,d^3\vec{p}= J^{ab}_i.
$$
Observe that Eq.\ref{eq19}  bears some resemblence to the continuity equation, 
except that it does not have 
the standard convective term. The absence may be attributed to the long 
wavelength approximation. The corresponding tensor charge would only be 
approximately conserved.

\subsubsection{ The $SU(3)$ gluons}
The generalization to the $SU(3)$ case is not difficult. The form of the 
distribution function is given by
\begin{equation}
f(x_i,p_i, \vec{Q})=f_0+\hat{Q} \cdot \vec{f}+\hat{Q}^{ab}f^{ab},
\end{equation} 
which is formally identical to Eq.\ref{eq17}. The transformation properties 
are however different: $f_0$ is the singlet, $f^a$ the octet, and the tensor 
$f^{ab}$ is  the 27-plet of $SU(3)$. The structure of 
the tensor basis $\hat{Q}^{ab}$ is more complicated than its counterpart 
in $SU(2)$ \cite{macfarlane}:

\begin{eqnarray}
\label{eq20}
 Q^2 \hat{Q}^{ab}=Q^aQ^b-\frac{3}{5}d^{pqr}d^{rab}Q^pQ^q-Q^2\frac{\delta^{ab}}{8},
\end{eqnarray}
where $d^{abc}$ are the symmetric structure constants of $SU(3)$. 
Note that $\hat{Q}^{ab}$ is symmetric and traceless. In all, there are $1+8+27$ 
parameters that characterize the color distribution, as it should be for 
a $8 \times 8$ density operator.
We shall not write the expressions for the Abelian and the non-Abelian 
permittivities since they are identical to Eq.\ref{eq17a} and Eq.\ref{eq17b}. 

 At this juncture we  note that there is an interesting relation between the 
permittivities of  quarks and gluons obeying the ideal equation of
state. Consider the susceptibilities 
${\cal A}(q,g) = \epsilon_A-1$ 
and ${\cal N}(q,g) = \epsilon_N-1$ 
for the quarks(q) and the gluons(g). 
It can be easily verified that the following relation holds 
\begin{eqnarray}
\label{eq22}
{\cal A}(q)=\frac{N_f}{2}{\cal A}(g),\nonumber\\
{\cal N}(q)=\frac{N_f}{2}{\cal N}(g).
\end{eqnarray}
It can be shown that it is true for any $SU(N)$ gauge group.

We turn our attention to the tensor fluctuations $f^{ab}$. They satisfy the 
eigenvalue equations 
\begin{equation}
R^{abcd}\phi^{cd}(\omega,k_i, p_i)
-2i L^{abcde} v_i \frac{1}{{\cal D}}{\cal A}^c_i \otimes \phi^{de} =0,
\end{equation}
where the rank four tensor, specific to SU(3) has the form 
$$
R^{abcd} = (\delta^{ac}\delta^{bd}+\delta^{ad}\delta^{bc}-\frac{6}{5}d^{abe}d^{ecd}),
$$
 and the rank five tensor has the form 
$$
 L^{abcde} =(f^{acd}\delta^{be}+f^{bcd}\delta^{ae}-\frac{6}{5}f^{mce}d^{abr}d^{rdm}).
$$
 A further integration over momentum yields, in the large wavelength limit,
\begin{equation}
\label{eq21}
R^{abcd}\rho^{cd}(\omega,k_i)
-2\frac{i}{\omega} L^{abcde} \frac{1}{{\cal D}}{\cal A}^c_i \otimes J_i^{de} =0
\end{equation}
which is similar to  Eq.\ref{eq19}. It has a more complicated structure with the
 occurrence of the symmetric as well as the antisymmetric structure constants of
the group. We study the above equations in a little bit more detail in the next
 section.
 
We wish to point out that one needs rather special fields to excite the tensor 
fluctuations. To demonstrate this aspect, let us consider Eq.\ref{eq21} when the
gauge fields represent a plane wave, i.e.,
$$
 A^a_i(\omega,k_i)=-i\frac{\vec{E}^a_{oi}}{\omega}\delta(\omega-\omega_0)\delta^3(k_i-k_{i0}).
$$
For this choice,  Eq.\ref{eq21} has only the trivial solution, {\it viz}, 
$f^{ab} \equiv 0$. One may expect that the 
tensor fluctuations may be excited provided the field has a quadrupole component
for $SU(2)$, and its generalization thereof for $SU(3)$. The implication of 
this result to the signatures of QGP needs to be explored.

\subsubsection{Closure in the color space}

We have seen that the color degree of freedom  does not possess an infinite 
hierarchy of moments since its associated phase space is compact. For $SU(2)$, 
the hierarchy of the moment equations terminates at the dipole level for the 
matter sector, and at the quadrupole level for the gluonic sector, as may be 
seen below:

\begin{widetext}
\begin{equation}
\label{eq18a}
{\phi}^b + iQ \frac{1}{{\cal D}}{{\cal E}}_i^b \frac{\partial f_0}{\partial p_i}
-i v_i\frac{1}{{\cal D}} (\vec{{\cal A}}_i \star \vec{{\cal \phi}})^b +\frac{3iV_2}{V_1}
\frac{1}{{\cal D}} {\cal E}_i^a \otimes \frac{ \partial \phi^{ab}}{\partial p_i}=0,
\end{equation}
\end{widetext}
 where $V_1, V_2$ are  defined in the Appendix. Similarly, the tensor 
fluctuations satisfy

\begin{widetext}
\begin{equation}
\label{eq18b}
\phi^{ab}+iQ \frac{1}{{\cal D}} \{{\cal E}^a_i \otimes \frac{\partial \phi^b}{\partial p_i} +
{\cal E}^b_i \otimes \frac{\partial \phi^a}{\partial p_i} \} -iL^{abcde} \frac{v_i}{{\cal D}}
{\cal A}^c_i \otimes \phi^{de} =0.
\end{equation}
\end{widetext}
The above equations are written without making the long wavelength 
approximation. This closure as exhibited in Eqs.\ref{eq18a}, \ref{eq18b} is 
formal since the 
gauge fields have to be determined self consistently with the matter 
distribution, as governed by the Yang-Mills equations. 

The corresponding equations for $SU(3)$ are slightly more complicated:

\begin{widetext}
\begin{equation}
{\phi}^b + iQ \frac{1}{{\cal D}}{{\cal E}}_i^b \frac{\partial f_0}{\partial p_i}
-i v_i\frac{1}{{\cal D}} (\vec{{\cal A}}_i \star \vec{{\cal \phi}})^b +\frac{16iV_2}{V_1}
\frac{1}{{\cal D}} {\cal E}_i^a \otimes T^{abcd} \frac{ \partial \phi^{cd}}{\partial p_i}=0
\end{equation}
where the tensor 
$$
T^{abcd} = \delta^{ac}\delta^{bd} - \frac{3}{5} d^{abe}d^{ecd}.
$$
The tensor excitations satisfy the equation
\begin{equation}
\label{eq21c}
R^{abcd}\big(\phi^{cd}+iQ \frac{1}{{\cal D}} \{{\cal E}^c_i \otimes \frac{\partial \phi^d}{\partial p_i} +
{\cal E}^d_i \otimes \frac{\partial \phi^c}{\partial p_i} \}\big) -iL^{abcde} \frac{v_i}{{\cal D}}
{\cal A}^c_i \otimes \phi^{de} =0
\end{equation}
\end{widetext}
where the two constants $V_1$ and $V_2$ are defined in appendix A.
We see sagin that the above two equations constitute a closed set, in a 
formal sense. Although their full solution in a self consistent manner is an 
intricate and an involved exercise, the clear signal from the above equations 
is that generically, all the fluctuations exist and none of the responses can 
be ignored. It also shows that if required, it is possible to improve upon our 
long wavelength approximation in a systematic manner. 

\section{ Results and discussion}

\subsection{\bf The Abelian response} 
Our analysis has made no assumption on the equilibrium distribution, save its 
isotropy in the momentum space, and independence of the position coordinates. 
We now consider, purely for purposes of illustration, an ideal gas of quarks 
and gluons. Of course, the Abelian component of the chromoelectric response 
is no different from its electrodynamic counterpart, except for color-spin 
factors. We merely record two of its properties.

Considering quarks first, we get
\begin{eqnarray}
\label{idab}
\lim_{\omega \rightarrow 0} \epsilon^A(\omega, k) & = & (1+\frac{m_D^2}{k^2}) \\
\lim_{k \rightarrow 0} \epsilon^A(\omega, k) & = & (1-\frac{m_D^2}{3\omega^2})
\end{eqnarray}
written in terms of the standard Debye mass 
$$m_D^2=4\pi^3Q^2T^2N_f/3.
$$
 The corresponding expressions for the gluonic plasma is obtained by simply 
replacing $N_F$ by 2 in the expression for $m_D^2$. It should be borne in mind 
that since we either have a gluonic plasma or a quark gluon plasma,
the expression for $m_D^2$ for QGP is obtained by averaging over the 
contributions from the quarks and gluons, weighted by the respective relative 
densities.

It is commonly accepted that the inverse Debye mass yields the screening 
length for the heavy quark potentials. This is so for electrodynamic 
plasmas, governed by the Coulomb potential, which goes
over to Yukawa. The situation is not that simple in the case of quarkonia. 
Since $J/\Psi$ suppression is one of the strong signatures for QGP, it is 
worth analyzing the import of the Debye mass to realistic
heavy quark potentials. We take it up in the next subsection.

\subsection{\bf The heavy quark potential}
Consider the Cornell potential given by \cite{char-data}
\[ \phi(r)=-\frac{\alpha}{r}+\Lambda r,\]
where $\alpha$ and $\Lambda$ are phenomenological constants. This potential 
has been studied in depth by Brambilla {\it et al} \cite{new-data}. The 
potential, has a Coulombic behavior at 
small distances and a linearly confining form at large distances. Note that 
the parameter $\Lambda$ has mass dimension 2 ($\alpha$ has mass dimension 0). 
The medium alters the above potential as given by the modified expression 
for the potential (in the Fourier space) 
$\phi(k) \rightarrow \phi_s(k) = \phi(k)/\epsilon(k)$.  $\phi_s(k)$ 
is obtained to be 
\begin{eqnarray}
\label{eq24}
\phi_s(k)=-\sqrt{\frac{2}{\pi}}\frac{\alpha}{k^2+m^2_D}
-\frac{4}{\sqrt{2\pi}}\frac{\Lambda }{k^2(k^2+m^2_D)}
\end{eqnarray}
The inverse Fourier transform of the above expression yields the modified 
potential, as a function of distance, as shown below:
\begin{eqnarray}
\label{eq27}
  \phi_s(r) = (\frac{2\Lambda}{m^2_D}-\alpha)\frac{\exp{(-m_D r)}}{r}-\frac{2\Lambda}{m^2_D r}.
\end{eqnarray}
The above expression merits a closer scrutiny. Interestingly, the Coulomb term 
goes over to the short range Yukawa form while the linear confining term 
goes over to a sum of Yukawa and the long range Coulomb potential. The 
Coulombic tail, which is relevant at large distances, supports -- as is 
universally 
known -- an infinite number of bound states. The effective Coulombic charge is  
 given by $\frac{2\Lambda}{m^2_D}$; this suggests that the role played by the 
Debye mass cannot be naively reduced to the notion of 
\noindent a screening length. Consequently, the relevant parameter to study the 
deconfinement is, {\it not the inverse Debye mass}, but the dissociation 
energy as a function of the parameter $\frac{2\Lambda}{m^2_D}$. Note that 
it is also the relevant parameter at high temperature.

Thus in the large distance/high temperature limit, it is not difficult to 
estimate the dissociation energy ${\cal E}_D$. We obtain
\begin{eqnarray}
{\cal E}_D \sim m_q\Lambda^2/m_D^4 \equiv \frac{3^2}{(4\pi^3)^2} \frac{m_q\Lambda^2}{Q^4 T^4N_F^2}.
\end{eqnarray}
It is further possible to estimate the temperature $T_D$ at which the 
dissociation takes place. Employing the equipartition theorem, we obtain
\begin{eqnarray}
T_D^5 = \frac{6}{(4 \pi^3)^2}\frac{m_q \Lambda^2}{ N_F Q^2}.
\end{eqnarray}
We note that for a purely gluonic plasma, the factor $N_F$ gets replaced by 2.

In short, a rather straightforward and a simple analysis has allowed us to 
extract the precise physical significance of the Debye mass for the 
dissociation, or equivalently, the suppression of the charmonium states in QGP. 
It also demonstrates that care should be exercised in interpreting the inverse 
mass as a screening length. The physics is dictated by a combination of two 
scales, $\Lambda$ and $m_D^2$, as obtained in Eq.\ref{eq27}. 

\subsection {\bf The non-Abelian response}
 We turn our attention to the non-Abelian permittivity. 
In the long wavelength limit, it has a simple expression given by 
\[\epsilon_N=1-\frac{m^2_D}{3\omega\omega^{\prime}}.\]
It is interesting to note that the above expression is almost 
identical to the expression for the Abelian permittivity at $k=0$. 
In fact, the quantity ( $\omega^2$ in $\epsilon_A$ 
gets simply replaced by $\omega\omega^{\prime}$ in $\epsilon_N$. 
The non-Abelian response vanishes at large 
$\omega,\omega^{\prime}$, and the response gets stronger with temperature. 
Of course, it vanishes when $T=0$, but it cannot be taken seriously since 
the interaction effects would dominate at small temperatures.
As mentioned, we may expect the non-Abelian response to play significant 
role in processes involving three gluon vertices.

\section{Landau damping in QGP}
 Landau damping in QGP has been studied earlier by Heinz and Siemens 
\cite{heinz-2} and Murtaza, Khattak and Shah \cite{murtaza}. We obtain here 
an explicit expression for Landau damping in QGP, arising especially from the 
non-Abelian response. A naive extension of the standard formula will not do 
since we have additional permittivities. Further the Abelian and non-Abelian 
components of the permittivities are not independent because of gauge 
invariance. We address the problem {\it ab initio}. 

The interaction Hamiltonian is given by
\begin{eqnarray}
\label{eq33}
 {\cal{H}}=J^a_\mu A^{a\mu}.
\end{eqnarray}
The energy dissipation per unit time per unit volume has the standard expression
\[ {\cal Q}= -\vec{J}_i\cdot \vec{E}_i.\]
Recall that we are in the temporal gauge. Employing Eq.\ref{currentfin}, 
which we reproduce below,
$$
\vec{J}_i = -Q^2 \vec{{\cal E}}_j \int d^3p v_i\frac{1}{{\cal D}} \frac{\partial f_0}{\partial_{p_j}},
$$
it is straight forward to obtain the contribution to the energy density due to 
the non-Abelian response. Finally, adding the standard Abelian contribution, 
the total energy density per unit time of an electric field is obtained to be
\begin{eqnarray}
\label{eq34}
 {\cal{U}}& =&\omega \{1+Q^2I_0(\omega,k)\}\vec{E}_i\cdot \vec{E}_i \nonumber \\  
 & & +Q^2I_1(\omega,k)|_{k=0}\vec{E}_i\cdot\vec{E}_i.
\end{eqnarray}
In Eq.\ref{eq34} the contribution from imaginary term gives the rate of 
the energy dissipation rate per unit volume, ${\cal E}_{dis}$. The 
contributions from $I_0$ and $I_1$ are respectively from the Abelian and 
the non-Abelian permittivities. 
 It is noteworthy that in the collisionless limit that we are interested in, 
only the Abelian component contributes to the damping, for, Eq.\ref{eq34} 
leads to the explicit form
\begin{eqnarray}
{\cal{U}}_{dis}=\pi m^2_D\{\frac{\omega^2}{2k^3}\theta(k-\omega)+\frac{\delta(\omega)}{3}\}\vec{E}_i\cdot\vec{E}_i.
\end{eqnarray}
The non-Abelian contribution is saturated at $\omega=0$ and is hence not 
an observable.

\section{Conclusion}
In conclusion, we have developed a systematic formulation for determining the 
chromoelectric response of QGP within the classical framework. We have shown
the emergence of a non-Abelian component of the response, characterized by
new permittivity $\epsilon_N$. $\epsilon_N$ is, unlike its 
Abelian component $\epsilon_A$, nonlocal and non-Markovian. 
Furthermore we have also derived explicitly the response functions of gluons 
which are shown to be richer than that of the quarks. 
The precise role of the Debye mass in the so called screening of the heavy 
quark potential is extracted. Nevertheless, we point out that the 
applications discussed in this paper are only indicative 
in nature since we have ignored the strongly interacting nature of QGP; 
our expressions for the equilibrium distribution 
functions are highly approximate. An application of the analysis made in 
this paper, with realistic equations of state as applied to QGP will be 
taken up separately. 

\section{Appendix}
\subsection{Color space integrals}

The Haar measure for the SU(3) group is given by
 \[ dQ=d^8Q\delta(Q^aQ_a-q^2)\delta(d^{abc}Q^aQ^bQ^c-q^3)\] \cite{heinz-1},
where $q^2=N$ and $q^3=0$ in the adjoint representation of SU(N) group 
\cite{cristina}.
Some useful color integrations in SU(3) group space are given by 
$$
\label{eq38b}
\noindent \int Q^a\,dQ=0; ~ \int Q^{ab}\,dQ=0
$$
$$
 \int Q^cQ^{ab}\,dQ=0; ~~
\int Q^aQ^b\,dQ=\frac{\delta^{ab}}{8}Q^2V_1;
$$
$$ 
\int Q^aQ^bQ^c\,dQ=0
$$  
$$
 \int Q^aQ^bQ^cQ^d\,dQ=V_2Q^4(\delta^{ab}\delta^{cd}+\delta^{ac}\delta^{bd}+\delta^{ad}\delta^{bc}).
$$
where $V_1$ is the volume of color space and $V_2$ is a constant. 
 $Q^{ab}$ has been defined as (see Eq.\ref{eq20})
\begin{eqnarray}
 Q^{ab}=Q^aQ^b-\frac{3}{5}d^{pqr}d^{rab}Q^pQ^q-\frac{\delta^{ab}}{8}Q^2.
\nonumber
\end{eqnarray} 



\begin{thebibliography} {05}

\bibitem{qgp-def}
 STAR collaboration, J. Adams {\it et al.}
Nucl. Phys. A {\bf 757} 102(2005).

\bibitem{exp-status1}
Itzhak Tserruya,
Nucl. Phys. A {\bf 774} 415(2006), Nucl.Phys.A {\bf 774} 433(2006).

\bibitem{exp-status2}
Patricia Fachini,
AIP Conf. Proc. {\bf 857} 62(2006).

\bibitem{s-a-bass}
 Steffan A. Bass,
Pramana {\bf 60} 593(2003). 

\bibitem{satz}
Helmut Satz,
Rep. Prog. Phys. {\bf 63} 1511(2000).

\bibitem{csorgo}
T. Cs\"{o}rg\"{o}, S. Hegyl, T. Nova\`{k} and W. A. Zajc,
Acta Phys. Polon. B {\bf 36} 329(2005). 

\bibitem{liq=he}
P. K. Kovtun, D.T. Son and A.O. Starinets,
Phys. Rev. Lett. {\bf 94} 111601(2005).

\bibitem{new-matter}
M. J. Tannenbaum,
Rept. Prog. Phys. {\bf 69} 2005(2006). 

\bibitem{lattice}
Frithjof Karsch,
Nucl. Phys. A {\bf 698} 199(2002).

\bibitem{lattice-new}
Y. Aoki, Z. Fodor, S. D. Katz and K. K. Szabo,
Proceedings of PoS LAT2005 163(2006). 

\bibitem{g-6}
K. Kajantie, M. Laine, K. Rummukainen and Y. Schroder,
Phys. Rev. D {\bf 67} 105008(2003).

\bibitem{fluid}
Ulrich Heinz,
Proceedings of *Swansea 2005, Extreme QCD* 3-12.

\bibitem{pisarski}
R. D. Pisarski,
Phys. Rev. Lett. {\bf 63} 1129(1989).

\bibitem{braaten-1}
E. Braaten and R. D. Pisarski,
Nucl. Phys. B {\bf 337} 569(1990).

\bibitem{braaten-2}
E. Braaten and R. D. Pisarski,
Nucl. Phys. B {\bf 339} 310(1990).

\bibitem{blaizot-1}
J. P. Blaizot and E. Iancu,
Nucl. Phys. B {\bf 417} 608(1994). 

\bibitem{blaizot-2}
J. P. Blaizot and E. Iancu,
Phys. Rev. Lett. {\bf 72} 3317(1994).

\bibitem{blaizot-3}
J. P. Blaizot and E. Iancu,
Nucl. Phys. B {\bf 421} 565(1994).

\bibitem{geiger}
K. Geiger,
Phys. Rep. {\bf 258} 237(1995).

\bibitem{lary}
A. Kovner, L. McLerran and H. Weigert,
Phys.Rev. D {\bf 52} 3809(1995).

\bibitem{ravi}
Gouranga C Nayak and V. Ravishankar,
Phys. Rev. C {\bf 58} 356 (1998); Phys. Rev. D {\bf 55} 6877 (1997).

\bibitem{raju}
F. Gelis and R. Venugopalan,
Lectures given at Cracow School of Theoretical Physics: 46th Course 2006, 
Zakopane, Poland, 27 May - 6 Jun 2006,
hep-ph/0611157.

\bibitem{bodeker}
D. Bodeker,
Phys. Lett. B {\bf 559} 502(1999).

\bibitem{matsui}
 T. Matsui and H. Satz,
 Phys. Lett. B {\bf 178} 416(1986).

\bibitem{weldon}
H. A. Weldon,
Phys. Rev. D {\bf 26} 1394(1982).

\bibitem{thoma}
Munshi G. Mustafa, Markus H. Thoma and Purnendu Chakraborty,
Phys. Rev. C {\bf 71} 017901(2005).

\bibitem{shuryak}
Jinfeng Liao and E. V. Shuryak,
Nucl. Phys. A {\bf 775} 224(2006). 

\bibitem{peter}
 P$\grave{e}$ter Petreczky,
hep-ph/9907247.

\bibitem{chen}
Ji-sheng Chen, Jia-rong Li and Peng-fei Zhuang,
Int. J. Mod. Phys. A {\bf 17} 1435(2002).

\bibitem{cristina}
Daniel F. Litim and Cristina Manuel,
Phys. Rep. {\bf 364} 451(2002). 

\bibitem{ravishankar}
Ambar Jain and V. Ravishankar,
Phys. Rev. Lett. {\bf 91} 112301(2003).

\bibitem{bhal}
R. S. Bhalerao and V. Ravishankar,
Phys. Lett. {\bf B 409} 38 (1997)

\bibitem{schmidt}
S. M. Schmidt et.al., 
Phys. Rev. D {\bf 59} 094005 (1999)

\bibitem{gross}
P. L. Bhatnagar, M. Gross and M. Crook,
Phys. Rev. {\bf 94} 511 (1954)

\bibitem{heinz}
Ulrich Heinz,
Phys. Rev. Lett. {\bf 51} 351(1983).

\bibitem{elze}
H. T. Elze and Ulrich Heinz,
Phys. Rep. {\bf 183} 81(1989).

\bibitem{perel-1}
A. Perelomov,
Generalized Coherent States and Their Applications, Springer Verlag (Berlin,
 Germany) 1986.

\bibitem{byrd}
Mark Byrd,
physics/9708015.

\bibitem{arxiv}
Akhilesh Ranjan, thesis (2008) unpublished

\bibitem{wong}
 S. K. Wong,
Nuovo Cimento A {\bf 65} 689(1970).

\bibitem{Wu}
Tai Tsun Wu and  Chen Ning Yang 
 Phys.Rev. {\bf D13} 3233 (1976).

\bibitem{macfarlane}
A. J. Macfarlane, A. Sudbery and P. H. Weisz,
Commun. math. Phys. {\bf 11} 77(1968).

\bibitem{char-data}
E. Eichten, K. Gottfried, T. Kinoshita, K. D. Lane and T. M. Yan,
Phys. Rev. D {\bf 21} 203(1980).

\bibitem{new-data}
N. Brambilla {\it et al},
CERN Yellow Report, CERN-2005-005, Geneva: CERN, 2005.- 487 p.

\bibitem{heinz-2}
Ulrich Heinz and Philip J. Siemens,
Phys. Lett. B {\bf 158} 11(1985).

\bibitem{murtaza}
G. Murtaza, N. A. D. Khattak and H. A. Shah,
Phys. Rev. E {\bf 68} (2003) 066404.

\bibitem{heinz-1}
 Ulrich Heinz,
Annals of Physics, {\bf 161} 48(1985).

\end{thebibliography}
\end{document}